\documentclass[runningheads]{llncs}
\usepackage{verbatim}
\usepackage{graphicx}
\usepackage{bm,amsfonts,makecell,multirow,stfloats,booktabs,subfigure,amsmath}
\usepackage[normalem]{ulem}
\usepackage[misc]{ifsym}
\useunder{\uline}{\ul}{}
\title{A Vlogger-augmented Graph Neural Network Model for Micro-video Recommendation}
\titlerunning{A Vlogger-augmented GNN Model for Micro-video Recommendation}
\author{Weijiang Lai\inst{1,2} \and
Beihong Jin\inst{1,2} \Letter \and
Beibei Li\inst{3} \and
Yiyuan Zheng\inst{1,2} \and
Rui Zhao\inst{1,2}}
\authorrunning{W. Lai et al.}
\institute{State Key Laboratory of Computer Science, Institute of Software, Chinese Academy of Sciences, Beijing, China \and
University of Chinese Academy of Sciences, Beijing, China \and
College of Computer Science, Chongqing University, Chongqing, China \email{Beihong@iscas.ac.cn}}

\begin{document}

\maketitle

\begin{abstract}
Existing micro-video recommendation models exploit the interactions between users and micro-videos and/or multi-modal information of micro-videos to predict the next micro-video a user will watch, ignoring the information related to vloggers, i.e., the producers of micro-videos. However, in micro-video scenarios, vloggers play a significant role in user-video interactions, since vloggers generally focus on specific topics and users tend to follow the vloggers they are interested in. Therefore, in the paper, we propose a vlogger-augmented graph neural network model VA-GNN, which takes the effect of vloggers into consideration. Specifically, we construct a tripartite graph with users, micro-videos, and vloggers as nodes, capturing user preferences from different views, i.e., the video-view and the vlogger-view. Moreover, we conduct cross-view contrastive learning to keep the consistency between node embeddings from the two different views. Besides, when predicting the next user-video interaction, we adaptively combine the user preferences for a video itself and its vlogger. We conduct extensive experiments on two real-world datasets. The experimental results show that VA-GNN outperforms multiple existing GNN-based recommendation models.


\end{abstract}

\section{Introduction}
Micro-video streaming platforms are hubs for uploading and watching micro-videos. In recent years, micro-video apps such as TikTok, Kwai, etc. attract a huge number of users, who spend most of their spare time watching diversified micro-videos. This also promotes more people to become vloggers, producing and publishing more micro-videos. With the increase in the number of micro-videos, the micro-video recommendation in the app becomes indispensable to users. 

Currently, some deep neural network models have been proposed for the micro-video recommendation, including ranking models and recall models. A few of the ranking models leverage multi-modal information including visual, acoustic, and textual features to achieve the recommendation \cite{ALPINE2019,MMGCN-MM2019,UVCAN-WWW2019,MTIN-MM2020}, thus being costly in terms of time and computational power. On the other hand, existing models for recalling micro-videos are relatively few in number \cite{CMI-SIGIR2022,PDMRec-ECML2022}. What is worse, existing micro-video recommendation models only exploit interactions between users and micro-videos for modeling and totally ignore the vlogger information.

We argue that vloggers play important roles in micro-video recommendation. Firstly, the vlogger of a micro-video can be treated as an attribute of the micro-video, since the vlogger potentially reflects the style of the micro-video. Meanwhile, a vlogger can also be treated as an attribute of a user, since the vloggers a user follows reveal the user's preferences. Taken together, treating vloggers as the auxiliary information of users or videos cannot adequately reflect the dynamic relationships among users, vloggers, and micro-videos. Secondly, the “Follow” function permits a user to find micro-videos newly published by followed vloggers. Thus, users are more likely to interact with the videos published by vloggers they follow. These user-video interactions may result from either user preferences for videos themselves or user preferences for their vloggers, the latter factor is overlooked by existing recommendation models. 

For taking the effect of vloggers on recommendation into consideration, in the paper, we construct a heterogeneous graph with users, micro-videos, and vloggers as nodes, and propose a model named \textbf{VA-GNN} (\textbf{V}logger-\textbf{A}ugmented \textbf{G}raph \textbf{N}eural \textbf{N}etworks) to exploit the complex semantic relationships among users, vloggers, and micro-videos. VA-GNN learns node embeddings from two different views, and set up meta-paths to build the connection between individual views in the original heterogeneous graph, thus improving the performance of micro-videos recommendations.


Our contribution can be summarized as follows.
\begin{itemize}
\item[$ \bullet$ ] We model the relationships among users, micro-videos, and vloggers in a tripartite graph and capture and combine user preferences for micro-videos themselves and vloggers so as to take full advantage of user-video interactions, user-vlogger interactions, and vlogger-video publishing relationships.

\item[$ \bullet$ ] We generate two embeddings for each node in the graph, using embedding propagation over the video view and the vlogger view, as well as the propagation along meta-paths which are formed by random walk across two views. Moreover, we employ cross-view contrastive learning to keep consistency between the two embeddings of the same node.



\item[$ \bullet$ ] We conduct extensive experiments on two real-world datasets. The experimental results show VA-GNN outperforms the other five models, in terms of Recall and NDCG.
\end{itemize}


\section{Related Work}
Our work is related to the research in three topics: GNN-based recommendation, contrastive learning for recommendation, and micro-video recommendation. 

\subsubsection{GNN-based recommendation.} GNNs are the neural networks that capture the dependence in a graph via message passing between nodes of the graph. They have been adopted by some recommendation models to perform different recommendation tasks, including the item recommendation \cite{NGCF} \cite{LightGCN} \cite{GTN}, social recommendation \cite{MHCN2021}, session recommendation \cite{SR-GNN2019}, bundle recommendation \cite{BGCN2020}, and cross-domain recommendation \cite{PPGN2019}.


Taking the item recommendation as an example, NGCF \cite{NGCF} models the user-item interactions as a bipartite graph and performs graph convolutions to update embeddings of nodes. Then the learned embeddings of nodes are used for the item recommendation. Further, LightGCN \cite{LightGCN} is proposed as a lightweight version of NGCF, which removes the nonlinear activation function and the feature transformation matrix in NGCF but has better performance than NGCF. Moreover, GTN \cite{GTN} is also a GNN model on the user-item bipartite graph but it can identify the reliability of the interactions, thus improving the performance.

\subsubsection{Contrastive learning for recommendations.} Contrastive learning is a branch of self-supervised learning. It has obtained great achievements in the fields of computer vision \cite{SimCLR} and NLP \cite{SimCSE}, and also helps improve the performance of recommendation models \cite{CrossCBPR}. 

Some sequential recommendation models, e.g., CL4SRec \cite{CL4SRec} and DuoRec \cite{DuoRec}, have been combined with contrastive learning. Typically, they generate the augmented sequences for the original sequence and then design an auxiliary task to pull positive sequence pairs closer to each other and push negative sequence pairs away from each other.

Contrastive learning is also applied to GNN-based recommendation models. Some models, e.g., SGL\cite{SGL}
and PCRec \cite{PCRec}, design perturbations in the structure of the original graph to obtain augmented graphs, and then with the aid of any GNN-based encoder, the contrastive learning task will drive embeddings of the same node existing in augmented graphs closer. Some models, e.g., EGLN\cite{EGLN} and BiGI\cite{BiGI}, construct the contrastive loss to keep the consistency between the local and global graphs. Further, SimGCL\cite{SimGCL} adds uniform noise directly to the embeddings of nodes in the graph, which develops a new way to obtain self-supervised signals.

\subsubsection{Micro-video recommendation.} Existing micro-video recommendation models have many model structures, including attention-based structures, CNNs, or GNNs. For example, Wei et al. \cite{MMGCN-MM2019} construct a user-item bipartite graph for each modality and generate modal-specific representations of users and micro-videos.

Some micro-video recommendation models have supplemented the contrastive learning component in their models. For example, CMI \cite{CMI-SIGIR2022} learns user multi-interests in micro-videos from historical interaction sequences and proposes a contrastive multi-interest loss to minimize the difference between interests extracted from two augmented views of the same interaction sequence. PDMRec \cite{PDMRec-ECML2022} applies a multi-head self-attention mechanism to learn sequence embeddings and proposes contrastive learning strategies to reduce the interference from micro-video positions in interaction sequences.

Compared to the existing work, our work models the relationship among users, micro-videos, and vloggers in a heterogeneous graph, and combines GNNs with contrastive learning for micro-video recommendation for the first time.

\section{Problem Formulation}

We use $ \mathcal{U}, \mathcal{V}, \mathcal{P}$ to denote the set of users, micro-videos, and vloggers, and define the user-video interaction matrix, user-vlogger interaction matrix, and vlogger-video publishing matrix as $ X_{|\mathcal{U}| \times|\mathcal{V}|}=\left\{x_{u v} \mid u \in \mathcal{U}, v \in \mathcal{V}\right\}$, $ Y_{|\mathcal{U}| \times|\mathcal{P}|}=\left\{y_{u p} \mid u \in \mathcal{U}, \notag\right.  \left. p \in \mathcal{P}\right\}$  and $ Z_{|\mathcal{P}| \times|\mathcal{V}|}=$  $ \left\{z_{p v} \mid p \in \mathcal{P}, v \in \mathcal{V}\right\}$ , respectively. If there is at least one explicit interaction (e.g., liking/thumb-up) between user $ u$  and micro-video $v$, then $ x_{u v}=1$, otherwise, $ x_{u v}=0$. Similarly, $ y_{u p}=1$  indicates that user $ u$  interacts with vlogger $ p$, for example, user $ u$ follows  vlogger $ p$. Moreover,  $ z_{p v} = 1$ indicates that vlogger $ p$  publishes video $v$. 

Our goal is to design a recommendation model to predict the probability of interaction between any user $ u \in \mathcal{U}$  and any candidate micro-video $ v \in \mathcal{V}$  based on the user-video interaction matrix $ X_{|\mathcal{U}| \times|\mathcal{V}|}$, user-vlogger interaction matrix $ Y_{|\mathcal{U}| \times|\mathcal{P}|}$  and vlogger-video publishing matrix $ Z_{|\mathcal{P}| \times|\mathcal{V}|}$.

\section{Our Model}
\begin{figure}[t]
\includegraphics[width=\textwidth]{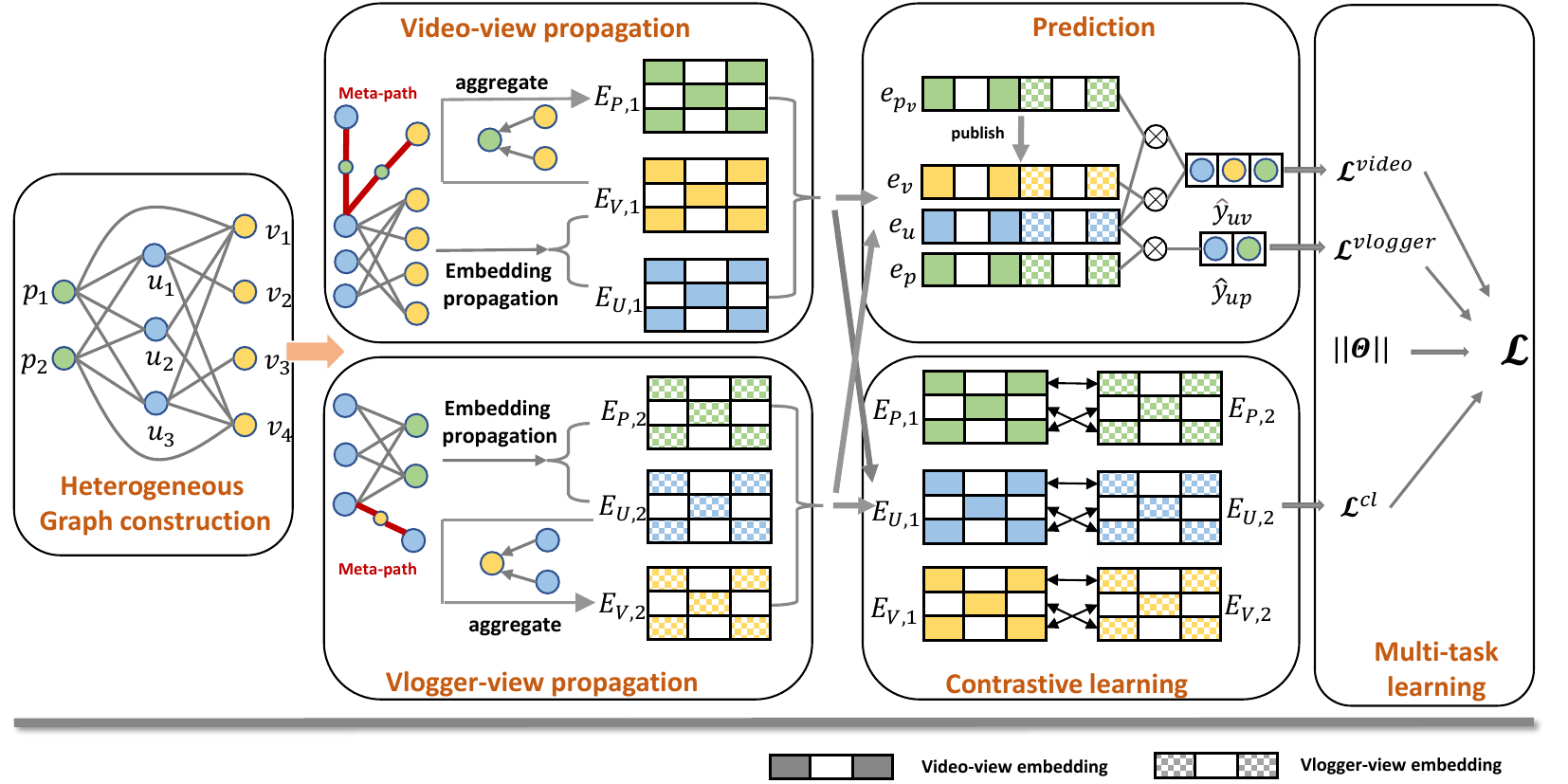}
\caption{Architecture of model VA-GNN.} \label{fig:model}
\end{figure}

We propose a model named VA-GNN, whose architecture is shown in Figure \ref{fig:model}. VA-GNN is composed of the following components: heterogeneous graph construction, embedding propagation, cross-view contrastive learning, prediction, and multi-task learning, which are detailed below.

\subsection{Heterogeneous Graph Construction}

To explicitly model the relationships between users, micro-videos, and vloggers, we construct a heterogeneous graph $ \mathcal{G}=(\mathcal{U}\cup\mathcal{V}\cup\mathcal{P}, \mathcal{E})$, where nodes consist of user nodes $ u \in \mathcal{U}$, micro-videos $ v \in \mathcal{V}$  and vlogger nodes $ p \in \mathcal{P}$, edges are $ \mathcal{E}$  consisting of user-video interaction edges $ (u, v)$  with $ x_{u v}=1$, user-vlogger interaction edges $ (u, p)$  with $ y_{u p}=1$, vlogger-video publishing edges $ (p, v)$  with $ z_{p v}=1$. 

We use $ \mathbf{e}_u \in \mathbf{E}_{\mathcal{U}}$, $ \mathbf{e}_v \in \mathbf{E}_{\mathcal{V}}$, $\mathbf{e}_p \in \mathbf{E}_{\mathcal{P}}$  to denote the embeddings of user $ u \in \mathcal{U}$, micro-video $ v \in \mathcal{V}$  and vlogger $ p \in \mathcal{P}$, respectively, where $\mathbf{E}_{\mathcal{U}}\in\mathbb{R}^{d\times{|\mathcal{U}|}}$, $\mathbf{E}_{\mathcal{V}}\in\mathbb{R}^{d\times{|\mathcal{V}|}}$, $\mathbf{E}_{\mathcal{P}}\in\mathbb{R}^{d\times{|\mathcal{P}|}}$, and $ d$  is the dimension of embeddings.

\subsection{Video-view Embedding Propagation}\label{sec:video-emb-proga}

In order to capture the high-order collaborative signals between users and micro-videos and learn more expressive user and micro-video embeddings, we exploit graph neural networks to propagate item-view information on the constructed heterogeneous graph based on user-video interactions. 

Meanwhile, we use the random walk to build the 'user-vlogger-user' meta-path and 'user-vlogger-video' meta-path to capture the impact of vlogger on users and micro-videos. Specifically, we construct a meta-path of length 3 by starting from a user $u$ and walking with probability $q_1$ to a vlogger $p$ interacted by $u$. Then we walk with probability $q_2$ to another user $u^{\prime}$ interacted by $p$ or $1-q_2$ to a video $v$ published by $p$, involving the nodes at the end of the meta-path in the embedding propagation of the item-view.
Here, we adopt a simple but effective embedding propagation operation, as shown follows.

\begin{equation}
\mathbf{e}_{u, 1}^{(l)}=\sum_{v \in \mathcal{V}_u} \frac{1}{\sqrt{\left|\mathcal{V}_u\right|} \sqrt{\left|\mathcal{U}_v\right|}} \mathbf{e}_{v, 1}^{(l-1)},    
\end{equation}

\begin{equation}
\mathbf{e}_{v, 1}^{(l)}=\sum_{u \in \mathcal{U}_v} \frac{1}{\sqrt{\left|\mathcal{U}_v\right| \sqrt{\left|\mathcal{V}_u\right|}}} \mathbf{e}_{u, 1}^{(l-1)},
\end{equation}

\begin{equation}
\mathbf{e}_{u, 1}^{(0)}=\mathbf{e}_u, \quad \mathbf{e}_{v, 1}^{(0)}=\mathbf{e}_p,    
\end{equation}
where $ \mathcal{V}_u$ denote micro-video neighbors and the 'user-vlogger-user' meta-path endpoints of user $ u$. $ \mathcal{U}_v$ denote user neighbors and the 'user-vlogger-video' meta-path endpoints of micro-video $ v$ .$ \mathbf{e}_{u, 1}^{(l)}$  and $ \mathbf{e}_{v, 1}^{(l)}$  denote the item-view embeddings of user $ u$  and micro-video $ v$  at the $ l$ -th propagation step.

For a vlogger, we aggregate the item-view embeddings of all the videos he/she published and obtain his/her item-view embedding, as follows.

\begin{equation}
\mathbf{e}_{p,1}^{(l)}=\mathtt{aggregate}\left(\mathbf{e}_{v,1}^{(l)} \mid v \in \mathcal{V}_p\right)    
\end{equation}
where $ \mathcal{V}_p$  denotes the micro-video neighbors of vlogger $ p$, i.e., the set of videos vlogger $ p$  published. Here, we directly take average pooling as the aggregator.

After iteratively performing embedding propagation for $ L$  steps, we obtain $ L$  embeddings for each user/video/vlogger node. We average the $ L$  embeddings of each node to generate its final item-view embedding, as follows.

\begin{equation}
\mathbf{e}_{u, 1}=\frac{1}{L} \sum_{l=0}^L \mathbf{e}_{u, 1}^{(l)}, \quad \mathbf{e}_{v, 1}=\frac{1}{L} \sum_{l=0}^L \mathbf{e}_{v, 1}^{(l)},\quad \mathbf{e}_{p, 1}=\frac{1}{L} \sum_{l=0}^L \mathbf{e}_{p, 1}^{(l)},
\end{equation}
where  $ \mathbf{e}_{u, 1}, \mathbf{e}_{v, 1}$, and $ \mathbf{e}_{p, 1}$  are item-view embeddings of user $ u$ , micro-video $ v$  and vlogger $ p$, respectively.

\subsection{Vlogger-view Embedding Propagation}

The videos published by the same vlogger usually fall in the same category. Thus, vlogger-view user preferences can reflect user preferences for certain video categories. For example, a vlogger who is a basketball coach mostly publishes videos about basketball skills, and users who follow the vlogger have a high probability of favoring basketball and are highly likely to be interested in other vloggers and micro-videos about basketball. 

Therefore, in order to learn the implied characteristics of each vlogger and capture vlogger-view user preferences, we perform vlogger-view embedding propagation based on user-vlogger interactions and calculate vlogger-view embeddings of users and vloggers. Similarly, we use the random walk to build the 'user-video-user' meta-path to capture the impact of micro-videos on users and vloggers,
involving the nodes at the end of the meta-path in the embedding propagation of the video-view. As shown follows.

\begin{equation}
\mathbf{e}_{u, 2}^{(l)}=\sum_{p \in \mathcal{P}_u} \frac{1}{\sqrt{\left|\mathcal{P}_u\right| \sqrt{\left|\mathcal{U}_p\right|}}} \mathbf{e}_{p, 2}^{(l-1)},
\end{equation}

\begin{equation}
\mathbf{e}_{p, 2}^{(l)}=\sum_{u \in \mathcal{U }_p} \frac{1}{\sqrt{\left|\mathcal{U}_p\right|} \sqrt{\left|\mathcal{P}_u\right|}} \mathbf{e}_{u, 2}^{(l-1)},
\end{equation}

\begin{equation}
\mathbf{e}_{u, 2}^{(0)}=\mathbf{e}_u, \mathbf{e}_{p, 2}^{(0)}=\mathbf{e}_p,
\end{equation}
where  $ \mathbf{e}_{u, 2}^{(l)}$  and $ \mathbf{e}_{p, 2}^{(l)}$  denote the vlogger-view embeddings of user $ u$  and vlogger $ p$ in the $ l$ -th layer, respectively. $ \mathcal{P}_u$ denote the neighbor vlogger nodes of user $ u$ . $ \mathcal{U}_p$ denote the neighbor user nodes and the 'user-video-user' meta-path endpoints of vlogger $ p$.

We calculate vlogger-view embeddings of micro-videos by aggregating the vlogger-view embeddings of their user neighbors. For example, the vlogger-view embedding of micro-video $ v$  is calculated as Equation \ref{eq:e_v2k}. For simplicity, we employ mean pooling as the aggregation function.

\begin{equation}\label{eq:e_v2k}
\mathbf{e}_{v, 2}^{(l)}=\mathtt{aggregate}\left(\mathbf{e}_{u, 2}^{(l)} \mid u \in \mathcal{U}_v\right)    
\end{equation}

 Similar to the final video-view embedding calculation in Section \ref{sec:video-emb-proga}, we take the average of the $ L$  vlogger-view embeddings of each node as the final vlogger-view embedding, as shown as follows.

\begin{equation}\label{eq:e_u2}
\mathbf{e}_{u, 2}=\frac{1}{L} \sum_{l=0}^L \mathbf{e}_{u, 2}^{(l)}, \quad \mathbf{e}_{v, 2}=\frac{1}{L} \sum_{l=0}^L \mathbf{e}_{v, 2}^{(l)},\quad \mathbf{e}_{p, 2}=\frac{1}{L} \sum_{l=0}^L \mathbf{e}_{p, 2}^{(l)}.
\end{equation}

\subsection{Cross-view Contrastive Learning}

After performing video-view and vlogger-view embedding propagation, we obtain pairwise embeddings from two different views for each user, micro-video, and vlogger, which are supposed to keep consistency. For example, user embeddings at video-view and vlogger-view both imply user preferences. 


Here, we facilitate embedding learning by conducting contrastive learning that mines the consistency between embeddings from different views of the same entity. Specifically, we treat the video-view and vlogger-view embeddings of the same user/video/vlogger as a positive pair, the embeddings of different users/videos/vloggers as negative pairs, and construct a cross-view contrastive learning loss as follows.
\begin{equation}
\mathcal{L}_{user}^{cl}=\frac{1}{|\mathcal{U}|} \sum_{u \in \mathcal{U}}-\log \frac{\exp \left(\mathtt{sim}\left(\mathbf{e}_{u, 1}, \mathbf{e}_{u, 2}\right) / \tau\right)}{\sum_{u^\prime \in \mathcal{U}} \exp \left(\mathtt{sim}\left(\mathbf{e}_{u, 1}, \mathbf{e}_{u^\prime, 2}\right) / \tau\right)},
\end{equation}

\begin{equation}
\mathcal{L}_{video}^{cl}=\frac{1}{|\mathcal{V}|} \sum_{v \in \mathcal{V}}-\log \frac{\exp \left(\mathtt{sim}\left(\mathbf{e}_{v, 1}, \mathbf{e}_{v, 2}\right) / \tau\right)}{\sum_{v^\prime \in \mathcal{V}} \exp \left(\mathtt{sim}\left(\mathbf{e}_{v, 1}, \mathbf{e}_{v^\prime, 2}\right) / \tau\right)},
\end{equation}

\begin{equation}
\mathcal{L}_{vlogger}^{cl}=\frac{1}{|\mathcal{P}|} \sum_{p \in \mathcal{P}}-\log \frac{\exp \left(\mathtt{sim}\left(\mathbf{e}_{p, 1}, \mathbf{e}_{p, 2}\right) / \tau\right)}{\sum_{p^\prime \in \mathcal{P}} \exp \left(\mathtt{sim}\left(\mathbf{e}_{p, 1}, \mathbf{e}_{p^\prime, 2}\right) / \tau\right)},
\end{equation}
where $ \mathtt{sim}(\cdot)$  is the cosine similarity function, $ \tau$  is the temperature parameter, which is a hyper-parameter.

The final contrastive loss is obtained by calculating the average of the three contrastive learning losses, as follows.

\begin{equation}
\mathcal{L}^{cl}=\frac{1}{3}\left(\mathcal{L}_{user}^{cl}+\mathcal{L}_{video}^{cl}+\mathcal{L}_{vlogger}^{cl}\right).
\end{equation}

\subsection{Prediction}

We concatenate user embeddings from two different views and obtain the final embedding of user $ u$  as $ \mathbf{e}_u=\mathbf{e}_{u, 1} \| \mathbf{e}_{u, 2}$, where $ \|$  denotes vector concatenation and $\mathbf{e}_u\in\mathbb{R}^{2d}$. In the same way, final embeddings of micro-video $ v$  and vlogger $ p$  are $ \mathbf{e}_v=\mathbf{e}_{v, 1} \| \mathbf{e}_{v, 2}$  and $ \mathbf{e}_p=\mathbf{e}_{p, 1} \| \mathbf{e}_{p, 2}$ , respectively. The predicted preference score between user $ u$  and vlogger $ p$  can be calculated as follows. 

\begin{equation}
\hat{y}_{u p}=\mathbf{e}_u^T \mathbf{e}_p.    
\end{equation}

Despite the target being micro-video itself, user preferences for vloggers who publish the video also play important roles in user-video interaction prediction. For example, in a situation where the target video does not fit well with the video-view preferences of a user, the user is still highly likely to interact with the video if he/she already follows the vlogger who published the video. Therefore, it is necessary to take user preferences for the vlogger of the target video into consideration when predicting the interaction score. Thus, we combine the user preference score for the video itself and the user preference score for its vlogger to obtain the final user-video interaction score as follows.

\begin{equation}
\hat{y}_{u v}=w \times \mathbf{e}_u^T \mathbf{e}_v+(1-w) \times \mathbf{e}_u^T \mathbf{e}_{p_v} ,
\end{equation}
where $ w\in(0, 1)$  is the weight of user preferences for the video $ v$  itself, $ p_v$  denotes the vlogger who published the video $ v$.

It is reasonable that if the content of a video is highly consistent with the focus of its vlogger, the video itself and its vlogger play comparable roles in interaction prediction since users like/dislike the video and its vlogger at the same time. But if the two are somewhat divergent, either the video itself or its vlogger plays a more important role. Motivated by that, we build a gate based on the correlation between the video and its vlogger to adaptively calculate weights of preferences at two different views as follows.  

\begin{equation}
    w=\sigma(\mathbf{e}_v^T Q \mathbf{e}_{p_v}),
\end{equation}
where $ \mathrm{Q} \in \mathbb{R}^{2d\times 2d}$  is model parameter to be trained. 

\subsection{Multi-task Learning}

We train our model by optimizing the multiple tasks, i.e., a micro-video recommendation task, a vlogger recommendation task, and a contrastive learning task.

For the micro-video recommendation task, we construct a BPR (Bayesian Personalized Ranking) loss as follows.

\begin{equation}
\mathcal{L}^{ {video }}=-\sum_{(u, v, v^-) \in S} \ln \sigma\left(\hat{y}_{u v}-\hat{y}_{u v^-}\right),
\end{equation}
where $\sigma$ is the sigmoid function,  $ S=\left\{({u}, {v}, {v^-}) \mid({u}, {v}) \in \mathcal{Y}^{+},(u, v^-) \in \mathcal{Y}^{-}\right\}$  denote the pairwise training data with negative sampling, $ \mathcal{Y}^{+}$ and $ \mathcal{Y}^{-}$  denote observable and unobservable user-video interactions, respectively.

In order to make the most of user-vlogger interaction data and capture vlogger-view user preferences more accurately, we leverage the vlogger recommendation task as an auxiliary task to train the model. Similarly, the training data of user-vlogger interactions can be denoted as $ S^{\prime}=\left\{({u}, {p}, {p^\prime}) \mid({u}, {p}) \in \mathcal{Z}^{+},(u, p^\prime) \in \notag\right. \\ \left. \mathcal{Z}^{-}\right\}$, where $ \mathcal{Z}^{+}$  denotes the set of observable positive user-vlogger pairs, and $ \mathcal{Z}^{-}$ denotes the set of unobservable negative user-vlogger pairs, we construct vlogger recommendation loss as follows.

\begin{equation}
\mathcal{L}^{ {vlogger }}=-\sum_{(u, p, p^-) \in S^{\prime}} \ln \sigma\left(\hat{y}_{u p}-\hat{y}_{u p^-}\right).
\end{equation}

Finally, we train the model by multi-task learning, and the final loss function is as follows.

\begin{equation}
\mathcal{L}=\mathcal{L}^{ {video }}+\lambda_1 \mathcal{L}^{ {vlogger }}+\lambda_2 \mathcal{L}^{cl}+\lambda_3\|\Theta\|,
\end{equation}
where $ \lambda_1, \lambda_2, \lambda_3$  are hyper-parameters that balance each loss function, and $ \|\Theta\|$  is the regularization term of the model parameters.



\section{Experiments}

\subsection{Experimental Settings}

\subsubsection{Datasets.} We conduct experiments on two real-world datasets, i.e., one public dataset and one industrial dataset.

\begin{itemize}
\item[$\bullet$] \textbf{WeChat-Channels} : This dataset is released by WeChat Big Data Challenge 2021. The dataset contains 14-day user interactions from WeChat-Channels, a popular micro-video platform in China.

\item[$\bullet$] \textbf{TakaTak}: This dataset is collected from TakaTak, a micro-video streaming platform for Indian users. The dataset contains user behaviors in four weeks.
\end{itemize}

Both two datasets include user-video interactions, user-vlogger interactions, and vlogger-video publishing relationships. 

We preprocess the datasets to clean user-video and user-vlogger positive interactions. We define posting comment, reading comments, liking, sharing, and so on as explicit positive feedback to videos. For the WeChat-Channels dataset, only the interactions which indicate explicit positive feedback from a user or the watching loop greater than 1.5 or the watching time greater than 60 seconds are retained. For the TakaTak dataset, the interactions with explicit positive feedback or completion rate greater than 1.8 or watching time greater than 15 seconds are defined as positive interactions. We also remove users and micro-videos with less than 5 interactions and remove vloggers who publish less than 3 micro-videos. As a result, the ratio of positive user-video interactions to non-positive ones approximately is equal to 1:3 for both datasets.

As for user-vlogger interactions, we treat following and entering the homepage as explicit positive behaviors and define a user-vlogger interaction if there exists an explicit positive interaction between them or the user interacts with more than two micro-videos published by the vlogger. Further, we remove users and vloggers with less than 5 interactions and obtain the final user-vlogger interactions.

The statistics of the processed user-video interactions and user-vlogger interactions are shown in Table \ref{tab:datsets}.

\begin{table}
\centering
\caption{Statistics of the datasets}
\setlength\tabcolsep{4.2pt}
\begin{tabular}{ccccccc}

\hline
Dataset        & \#Users & \#Micro-videos & \#Vloggers & \#User-video & \#User-vlogger  \\ 
& & & & Interactions & Interactions  \\
\hline
WeChat         & 19739   & 25976   & 2088               & 1490633  & 34450\\
TakaTak  & 14571   & 13133   & 912             & 1661032  & 262548\\
\hline
\end{tabular}
\label{tab:datsets}
\end{table}




After sorting user interactions in ascending order by timestamp, we use the leave-one-out method to divide a dataset into train/validation/test sets. For each user, the last interaction is used for testing, the interaction before the last one is used for validation, and the remaining interactions are used for training.

\subsubsection{Metrics}
We adopt $Recall@K$ and $NDCG@K$ as metrics of performance evaluation, where Recall focuses on whether the recommended micro-videos are hit or not and NDCG focuses on the ranking of the recommended micro-videos. We set $K$ to 10, 20, and 50.

\subsubsection{Competitors}
To evaluate the performance of our model, we choose the following models as the competitors.

\begin{itemize}
\item[$\bullet$]\textbf{NGCF} \cite{NGCF}: a GNN-based recommendation model. Based on the idea of collaborative filtering, the model explicitly models the high-order connectivity between users and micro-videos through GNN, which is beneficial to embedding learning.

\item[$\bullet$]\textbf{LightGCN} \cite{LightGCN}: a GNN-based recommendation model. The model finds that the nonlinear activation function and feature transformation matrix in NGCF degrade the performance, and it only utilizes simple weighting and aggregation methods.

\item[$\bullet$]\textbf{GTN} \cite{GTN}: a GNN-based recommendation model. The model considers that not all user interactions are reliable and designs graph neural networks to capture interaction reliability.

\item[$\bullet$]\textbf{SGL} \cite{SGL}: a recommendation model that introduces contrastive learning to GNN. The model generates contrastive views by employing node drop (ND), edge drop (ED), or random walk (RW), and then adds a contrastive loss to align the embeddings of positive pairs.

\item[$\bullet$]\textbf{SimGCL} \cite{SimGCL}: a recommendation model that integrates with a  contrastive learning task. The model gives up the data augmentation at the graph level in SGL and proposes an augmentation method at the embedding level, that is, to construct positive pairs by adding uniform noise to the embeddings. 
\end{itemize}

\subsubsection{Implementation Details}

For all the competitors, we adopt the implementations of open-source code. The hyperparameters of competitors are tuned based on their original papers. For a fair comparison, we initialize all the model parameters with a normal distribution, set the embedding size to 64, and set the batch size to 4096.

Our model is implemented by PyTorch. We use a random negative sample method and set the number of negative samples to 1 for the BPR loss. We use Adam with a learning rate of 0.001 to optimize our model. For the other hyperparameters, we tune $ \lambda_1$, $ \lambda_2$  and $ \tau$  within \{0.01,0.05,…,5,10\}, \{0.0001,0.0005,…,1,5\} and \{0.01,0.05,…,1,5\}, respectively. The model reaches the optimal performance on the WeChat-Channels dataset with $ \lambda_1$, $ \lambda_2$, and $ \tau$ set to 5, 0.0005, and 0.5 and on the TakaTak dataset with $ \lambda_1$, $ \lambda_2$, and $ \tau$ set to 5, 0.05, and 0.05.

We train the model with an early stopping strategy. That is, we stop training the model if Recall@10 does not increase on the validation set for 10 epochs consecutively. We adopt model parameters achieving the best performance on the validation set for testing. Our model is not allowed to recommend micro-videos that have been watched by the user. Our implementation code is available at https://github.com/laiweijiang/VAGNN.git.

\subsection{Performance Comparison}

The performance of all models on two datasets is listed in Table~\ref{tab:performance}. From the results, we have the following observations.

\begin{table}[tb]
\centering
\caption{Recommendation performance on two datasets. }
\setlength\tabcolsep{4.2pt}

\footnotesize
\begin{tabular}{cccccccc}
WeChat-Channels  & & & & & & \\
\hline
\multicolumn{1}{c|}{ \multirow{2}{*}{Model}}         & \multicolumn{3}{c}{Recall}& \multicolumn{3}{c}{NDCG}\\ 
\multicolumn{1}{c|}{} & @10 & @20 & @50 & @10 & @20 & @50\\
 \hline
\multicolumn{1}{c|}{NGCF}  & 0.0382 & 0.0698 & 0.1356 & 0.0185 & 0.0264 & 0.0393 \\
\multicolumn{1}{c|}{LightGCN} & 0.0441 & 0.0742 & 0.1413 & 0.0221 & 0.0296 & 0.0428 \\
\multicolumn{1}{c|}{GTN} & 0.0433 & 0.0721 & 0.1385 & 0.0214 & 0.0286 & 0.0417 \\
\multicolumn{1}{c|}{SGL-ED} & 0.0451 & 0.0751 & 0.1454 & 0.0222 & 0.0297 & 0.0435 \\
\multicolumn{1}{c|}{SGL-ND} & 0.0441 & 0.0723 & 0.1404 & 0.0220 & 0.0291 & 0.0425 \\
\multicolumn{1}{c|}{SGL-RW} & 0.0452 & 0.0752 & 0.1451 & 0.0225 & 0.0300 & 0.0438 \\
\multicolumn{1}{c|}{SimGCL} & {\ul 0.0470} & {\ul 0.0801} & {\ul 0.1518} & {\ul 0.0233} & {\ul 0.0317} & {\ul 0.0458} \\
\multicolumn{1}{c|}{VA-GNN} & {\bfseries 0.0503} & {\bfseries 0.0826} & {\bfseries 0.1545} & {\bfseries 0.0248} & {\bfseries 0.0330} & {\bfseries 0.0471} \\
\hline
\multicolumn{1}{c|}{Improv. (\%)} & 7.02 & 3.12 & 1.78 & 6.44 & 4.10 & 2.84 \\
\hline

 \\
 TakaTak  & & & & & & \\
 \hline
\multicolumn{1}{c|}{\multirow{2}{*}{Model}}         & \multicolumn{3}{c}{Recall}& \multicolumn{3}{c}{NDCG}\\ 
\multicolumn{1}{c|}{} & @10 & @20 & @50 & @10 & @20 & @50\\
 \hline
\multicolumn{1}{c|}{NGCF} & 0.0437 & 0.0810 & 0.1769 & 0.0204 & 0.0298 & 0.0487 \\
\multicolumn{1}{c|}{LightGCN} & 0.0478 & 0.0867 & 0.1837 & 0.0227 & 0.0324 & 0.0514 \\
\multicolumn{1}{c|}{GTN} & 0.0461 & 0.0852 & 0.1779 & 0.0224 & 0.0321 & 0.0504 \\
\multicolumn{1}{c|}{SGL-ED} & 0.0481 & 0.0885 & 0.1809 & 0.0229 & 0.0330 & 0.0511 \\
\multicolumn{1}{c|}{SGL-ND} & {\ul 0.0497} & {\ul 0.0903} & {\ul 0.1840} & {\ul 0.0241} & {\ul 0.0341} & {\ul 0.0527} \\
\multicolumn{1}{c|}{SGL-RW} & 0.0467 & 0.0881 & 0.1833 & 0.0226 & 0.0329 & 0.0516 \\
\multicolumn{1}{c|}{SimGCL} & 0.0471 & 0.0863 & 0.1823 & 0.0229 & 0.0327 & 0.0516 \\
\multicolumn{1}{c|}{VA-GNN} & {\bfseries 0.0622} & {\bfseries 0.0987} & {\bfseries 0.1870} & {\bfseries 0.0320} & {\bfseries 0.0411} & {\bfseries 0.0585} \\
\hline
\multicolumn{1}{c|}{Improv. (\%)} & 25.15 & 9.30 & 1.63 & 32.78 & 20.53 & 11.01 \\
\hline
\end{tabular}
\label{tab:performance}
\end{table}

Our model achieves the best performance and outperforms all the competitors on all datasets in terms of all metrics, indicating the superiority of our model. We attribute this result to that we effectively model the complex relationship among users, micro-videos, and vloggers, and utilize contrastive learning to further enhance the performance of the model. LightGCN outperforms NGCF in all datasets, which is consistent with the claim in \cite{LightGCN}. However, the performance of the GTN model is lower than LightGCN, presumably because we removed users, micro-videos, and vloggers with few interactions during the data processing stage. It may also remove some noise data simultaneously, which is not beneficial for GTN.

Our model outperforms LightGCN by a large margin. For example, our model outperforms LightGCN by 14.06$ \%$  and 30.13$ \%$  on Recall@10 in WeChat-Channels and TakaTak datasets, respectively. Compared to LightGCN which conducts embedding propagation on the user-video bipartite graph, we perform embedding propagation on the user-video-vlogger tripartite graph. The experimental results show that vloggers do contain extensive information, such as the implied user preferences, in the micro-video scenario. Mining the information related to vloggers helps our model enhance the quality of embedding.

The three SGL variants and SimGCL outperform LightGCN in most metrics. Besides, SimGCL and SGL-ND are the second-best models on WeChat-Channels and TakaTak datasets, respectively, which demonstrates the effectiveness of contrastive learning in graph recommendation models. However, SimGCL only has the average performance on TakaTak, and also SGL-ND shows mediocre performance on WeChat-Channels. The reason might be that in the data augmentation phase, the approach of adding uniform noise by SimGCL and the approach of randomly dropping nodes and surrounding edges by SGL-ND greatly change the original embedding and graph structure, respectively, resulting in unstable performance on different datasets. On the contrary, VA-GNN adopts cross-view contrastive learning that directly relies on the structure of the graph, thus benefiting stably from contrastive learning.

\subsection{Ablation Study}

\begin{table}[tb]
\centering
\caption{Ablation study.}
\setlength\tabcolsep{4.2pt}
\footnotesize
\begin{tabular}{ccccccc}
WeChat-Channels  & & & & & & \\
\hline
 \multicolumn{1}{c|}{\multirow{2}{*}{Model}}          & \multicolumn{3}{c}{Recall}& \multicolumn{3}{c}{NDCG}\\ 
 \multicolumn{1}{c|}{}& @10 & @20 & @50 & @10 & @20 & @50\\
 \hline
       
\multicolumn{1}{l|}{VA-GNN} & \textbf{0.0503} & \textbf{0.0826}  & \textbf{0.1545}   & \textbf{0.0248}       & \textbf{0.0330}       & \textbf{0.0471}      \\
\multicolumn{1}{l|}{(A) w/o CL loss} & 0.0470 & 0.0778  & 0.1492   & 0.0235   & 0.0312 & 0.0453      \\
\multicolumn{1}{l|}{(B) w/o vlogger loss} & 0.0424 & 0.0732  & 0.1439   & 0.0214   & 0.0291      & 0.0430      \\
\multicolumn{1}{l|}{(C) w/o video-view} & 0.0240 & 0.0421  & 0.0921   & 0.0114       & 0.0159       & 0.0257      \\
\multicolumn{1}{l|}{(D) w/o vlogger-view} & 0.0443 & 0.0725  & 0.1313   & 0.0226       & 0.0296       & 0.0412      \\
\multicolumn{1}{l|}{(E) w/o $ \hat y_{up}$}  & 0.0477 & 0.0811  & 0.1518   & 0.0238       & 0.0322       & 0.0461      \\ \hline
\\
TakaTak  & & & & & & \\
\hline
 \multicolumn{1}{c|}{\multirow{2}{*}{Model}}          & \multicolumn{3}{c}{Recall}& \multicolumn{3}{c}{NDCG}\\ 
 \multicolumn{1}{c|}{}& @10 & @20 & @50 & @10 & @20 & @50\\
 \hline
\multicolumn{1}{l|}{VA-GNN} & \textbf{0.0622} & \textbf{0.0987}  & \textbf{0.1870}   & \textbf{0.0320}       & \textbf{0.0411}       & \textbf{0.0585}      \\
\multicolumn{1}{l|}{(A) w/o CL loss} & 0.0605 & 0.0980  & 0.1791   & 0.0303       & 0.0396       & 0.0555      \\
\multicolumn{1}{l|}{(B) w/o vlogger loss} & 0.0488 & 0.0838  & 0.1721   & 0.0226       & 0.0314       & 0.0487      \\
\multicolumn{1}{l|}{(C) w/o video-view} & 0.0349 & 0.0662  & 0.1371   & 0.0155       & 0.0233       & 0.0372      \\
\multicolumn{1}{l|}{(D) w/o vlogger-view} & 0.0465 & 0.0817  & 0.1662   & 0.0215       & 0.0303       & 0.0469      \\
\multicolumn{1}{l|}{(E) w/o $ \hat y_{up}$}  & 0.0463 & 0.0837  & 0.1727   & 0.0222       & 0.0315       & 0.0490      \\ \hline
\end{tabular}
\label{tab:ablation}
\end{table}
We evaluate the effectiveness of designed modules through the ablation study. We construct five variants as shown in Table \ref{tab:ablation}, where variant A is VA-GNN without contrastive loss. Variant B is VA-GNN without the vlogger recommendation loss. Variant C is VA-GNN without video-view embedding propagation. Variant D is VA-GNN without vlogger-view embedding propagation. The variant E is VA-GNN  that predicts interaction scores without considering user preferences for the vlogger of the target video, i.e., predicting the user-video interaction score between user $u$ and video $v$ as $ \hat{y}_{u v}=\mathbf{e}_u^T \mathbf{e}_v$. The results on two datasets are shown in Table~\ref{tab:ablation}. 

From the results, we find that variants C and D have a substantial decline in performance, compared to VA-GNN, indicating that embedding propagations over video-view and vlogger-view are critical in modeling and mining the relationship among users, micro-videos and vloggers. The performance of variant E is comparable to LightGCN but shows a significant decrease compared to VA-GNN. This indicates that user preferences for a micro-video originate from both the micro-video itself and the vlogger who publishes the video. Therefore, combining the user preference score for the video itself and the user preference score for its vlogger to predict the final user-video interaction score can effectively improve the model performance. Further, the performance of variant B shows that learning from positive user-vlogger pairs can substantially improve the model performance. Finally, the variant model (A) shows a slight performance decline, compared to VA-GNN, this indicates that contrastive learning can boost the performance of the model a bit.

\subsection{Hyperparameter Sensitivity Analysis}

\begin{figure}[t]
\includegraphics[width=\textwidth]{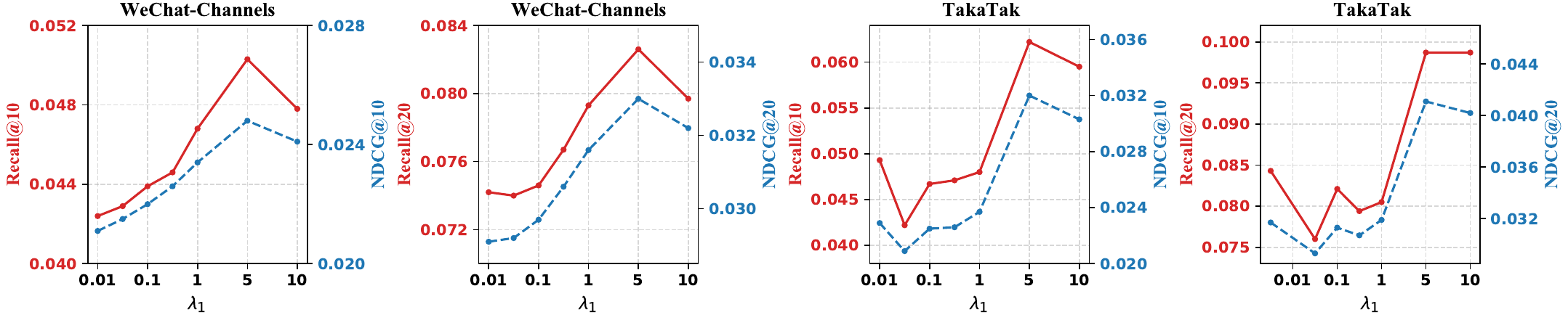}
\caption{ Sensitivity of vlogger loss weight $ \lambda_1$  on two datasets.} \label{tab:lmd1}
\end{figure}
\begin{figure}[t]
\includegraphics[width=\textwidth]{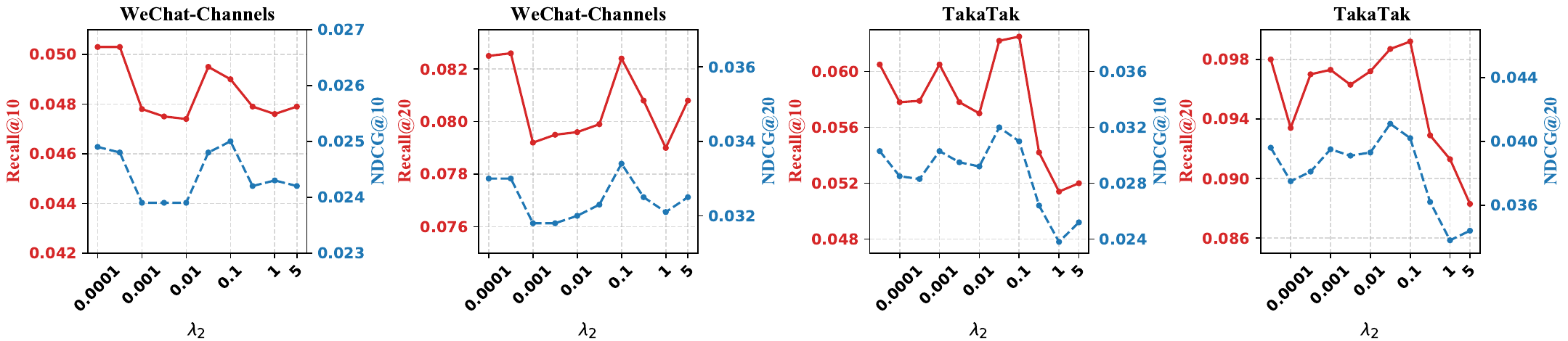}
\caption{ Sensitivity of contrastive loss weight $ \lambda_2$  on two datasets.} \label{tab:lmd2}
\end{figure}
\begin{figure}[tb]
\includegraphics[width=\textwidth]{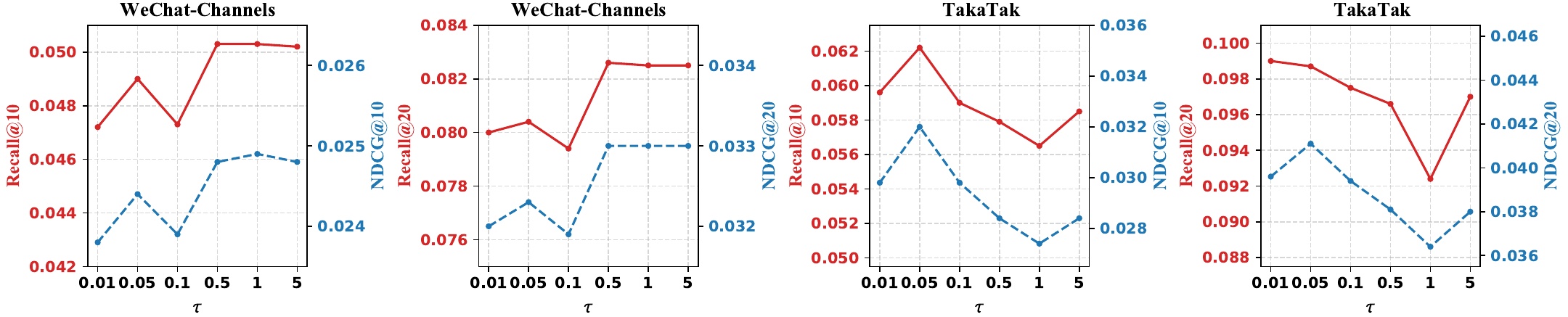}
\caption{ Sensitivity of the temperature $ \tau$  on two datasets.} \label{tab:tau}
\end{figure}

We conduct experiments on two datasets to observe the impact of different values of hyperparameters (i.e., vlogger loss weight $ \lambda_1$, contrastive loss weight $ \lambda_2$ and the temperature $ \tau$) on the performance. Figures~\ref{tab:lmd1}, \ref{tab:lmd2}, and \ref{tab:tau} show the performance changes with the change of these hyperparameters, respectively.

From Figure ~\ref{tab:lmd1}, we find that the model performance increases sharply when the value of $\lambda_1$ is set from 0.01 to 5 and achieves the maximum when $\lambda_1$ is 5, and then decreases slowly when $\lambda_1$ is set from 5 to 10. Obviously, choosing an appropriate $ \lambda_1$ has a large impact on the performance. From Figure~\ref{tab:lmd2}, we find the choice of $\lambda_2$ is directly related to the dataset. On the WeChat-Channels dataset, the model performs best with $ \lambda_2$ of 0.0005, on the WeChat-Channels dataset, the model performs best with $ \lambda_2$ of 0.1 on the TakaTak dataset.

The temperature controls the sensitivity of hard samples and the tolerance of similar samples. In general, a low temperature is beneficial for mining hard negative samples, but a quite low value may damage the semantic structure. From Figure~\ref{tab:tau}, we can find that a high or low $ \tau$ has different impacts on different datasets. Besides, TakaTak dataset is more sensitive to the temperature than WeChat-Channels dataset, and deviating from the best setting will degrade the performance remarkably. That is because the effect of temperature will be magnified by $\lambda_2$. $\lambda_2$ on TakaTak dataset needs to be set relatively large to achieve good performance. Therefore, changes in temperature on TakaTak dataset can obviously affect the performance.

\section{Conclusion}
The popularity of micro-video apps benefits from a large number of micro-videos produced by plenty of different vloggers. Inspired by this characteristic, VA-GNN models the relationship among users, micro-videos, and vloggers, and mines user preferences for micro-videos as well as vloggers for recommendations. VA-GNN also incorporates contrastive learning into graph neural networks, thus achieving the best micro-video recommendation performance on two real-world datasets while comparing to five existing models.

\section*{Acknowledgment}
This work was supported by the National Natural Science Foundation of China under Grant No. 62072450 and the 2021 joint project with MX Media.

\section*{Ethical Statements}

Hereby, we consciously assure that the data used in the experiments are desensitized and do not contain personal privacy.
%
%
%
\bibliographystyle{unsrt}
\bibliographystyle{splncs04}
\bibliography{mybibliography}
%





\end{document}